\documentclass[preprint,12pt]{elsarticle}

\usepackage{amssymb}
\journal{New Astronomy}

\begin{document}

\begin{frontmatter}



\title{On the propensity of the formation of massive clumps via fragmentation of driven shells}


\author[SVA]{Anathpindika, S.}
\ead{sumedh$\_$a@iiap.res.in}
\address{Indian Institute of Astrophysics, 2$^{nd}$-Block, Koramangala, Bangalore-560034, India}

\begin{abstract}
   Early type massive stars drive thin, dense shells whose edges often show evidence of star-formation. The possibility of fragmentation of these shells, leading to the formation of putative star-forming clumps is examined with the aid of semi-analytic arguments. We also derive a mass-spectrum for clumps condensing out of these shells by performing Monte-Carlo simulations of the problem. By extending on results from our previous work on the stability of thin, dense shells, we argue that clump-mass estimated by other authors in the past, under a set of simplifying assumptions, are several orders of magnitude smaller than those calculated here. Using the expression for the fastest growing unstable mode in a shock-confined shell, we show that fragmentation of a typical shell can produce clumps with a typical mass $\gtrsim 10^{3}$ M$_{\odot}$. It is likely that such clumps could spawn a second generation of massive and/or intermediate-mass stars which could in turn, trigger the next cycle of star-formation. We suggest that the ratio of shell thickness-to-radius evolves only weakly with time. Calculations have been performed for stars of seven spectral types, ranging from B1 to O5. We separately consider the stability of supernova remnants.
\end{abstract}

\begin{keyword}
star formation \sep HIIregions \sep supernova remnants \sep instabilities
\end{keyword}

\end{frontmatter}

% \linenumbers

\section{Introduction}
Partial or full ring-like structures, which apparently are projections of shells, are often found in the interstellar medium (ISM). These dense shells could possibly be driven by one of the numerous sources like, ionising radiation from young star-clusters, early type massive stars, blast-waves from supernovae, or energetic stellar winds. Detailed observations of these shells in various bands of the infrared wavelength have also revealed isolated sites of massive star formation (e.g. Deharveng, Zavagno \& Caplan 2005). In the recent past a number of such sites have been reported, for instance in the HII region RCW79 (Zavagno \emph{et al.} 2006), and RCW120 (Anderson \emph{et al.} 2010). A catalogue of 600 such shells in the galactic disk was drawn up by Churchwell \emph{et al.} (2006) as part of the GLIMPSE survey. The survey showed that a large proportion ($\sim$90 \%) of these shells are thin, i.e. the shell thickness is less than a third of the outer shell radius, and driven primarily by massive stars ($\sim86\%$ shells), see also Zavagno \emph{et al.} 2010. 

It is well-known that relatively high-mass stars emit powerful radiation that ionises gas in the local neighbourhood, and heats it to temperatures typically of the order of 10$^{4}$ Kelvin. This hot plasma propagates in the interstellar medium (ISM) at a highly supersonic speed (typical velocity of expansion in the initial phases is $\sim 10^{3}$ km/s), whence it gradually equilibrates to a significantly lower temperature, of order a few thousand Kelvin. The expanding, roughly spherical volume of hot plasma, the so called Stroemgren sphere, cools primarily via collisional excitation of heavier elements such as Carbon, nitrogen, and oxygen, and sputtering of dust-grains while sweeping up a dense shell of gas in the ISM; the familiar snow-plough phase. This shell is confined by two shocks, first, due to the wind driving it, and second, due to the reverse shock resulting from the propagation of the shell in the ISM. The stability of such shells has been discussed by numerous authors, e.g. Elmegreen \& Elmegreen (1978), Larson (1985), Elmegreen (1989), Vishniac (1983, 1994), Whitworth (1994), Wuensch \& Palou$\breve{s}$ (2001) and Anathpindika (2010). It has been demonstrated by these and several other authors that, a shock-confined shell is unstable to instabilities arising out of shock-induced turbulence within layers of the shell. Turbulence leads to enhanced transfer of momentum in different regions of the shell that makes it unstable to the so called thin shell instability (TSI) (Vishniac 1983, 1994), and raises the effective local sound-speed. The stability of the shell depends on the critical interplay between the gravitational instability (GI) and the TSI. 

A full fledged analytic treatment of the stability of a shock-confined slab is rather complex as has been demonstrated by Vishniac (1994), for instance, who showed that such slabs were likely to be unstable to the so called non-linear thin shell instability (NTSI). Traditionally, stability analysis of shocked shells and/or slabs, for all practical purposes, have been simplified by excising shock dynamics in favour of a simple high-pressure approximation thereby eliminating perturbative effects of the TSI, and its non-linear excursion (e.g. Whitworth \emph{et al.} 1994b). Anathpindika (2009, 2010) numerically showed that a shocked slab and/or shell is unstable to the TSI, which, soon after its formation, develops wiggles on its surface and grows non-linearly. Ehlerov$\grave{a}$ \& Palou$\breve{s}$ (2002) derived the critical density for a shell to become gravitationally unstable. Under a sinusoidal approximation for these perturbations, Anathpindika (2010) deduced an expression for the wavenumber of the fastest growing mode. Below, we propose to test the validity of this expression for HII shells driven by typical candidate stars listed in Table 5.3 of Spitzer (1978), and deduce a mass function for clumps condensing out of these shells. The case of a shell driven by a supernova blast wave will be considered separately. We shall demonstrate that the mass function so derived is consistent with that reported by Fukui \emph{et al.}(1999), Yamaguchi \emph{et al.} (2001), and Roslowsky (2005) for massive clouds. In \S 2 we shall deduce our set of equations and demonstrative calculations, including the calculation of a mass spectrum for fragments,  will be undertaken in \S 3. We conclude in \S 4.

\section{Semi-analytic deduction }
\subsection{HII shells}

 Let us consider a typical source of ionising radiation that emits N$_{LyC}$ number of photons per second. If $n_{p}$, $n_{e}$ are the respective number of protons and $e^{-}$ per unit volume, then $xn_{e}n_{p}\alpha^{(2)}$ is the number of electrons captured per cm$^{-3}$ in the ground state, and the flux of photons flowing through a shell of radius, $r_{S}$, is-
\begin{equation}
\frac{4\pi}{3}r_{S}^{3}xn_{e}n_{p}\alpha^{(2)} = \mathrm{N}_{LyC},
\end{equation}  
where $\alpha^{(2)}$ is the recombination coefficient that excludes electron captures to the ground state, and defined as 
\begin{displaymath}
\alpha^{(2)} = \frac{2.06\times 10^{-11}Z^{2}}{T^{1/2}}\phi_{2}(\beta) \mathrm{cm}^{-3}\mathrm{s}^{-1},
\end{displaymath}
at a temperature $T$; $Z$ is the ionic charge and $\phi_{2}(\beta)$ is the recombination coefficient function corresponding to $\alpha^{(2)}$ (see Table 5.2, Spitzer 1978). The ionisation fraction of hydrogen, $x$, has been set equal to unity so that the Stroemgren radius is defined as -
\begin{equation}
r_{s}^{3} = \frac{3\mathrm{N}_{LyC}}{4\pi n^{2}\alpha^{(2)}},
\end{equation}
where the condition of approximate charge neutrality forces $n_{e}\sim n_{p}\equiv n$. 

 The temporal evolution of the ionised shell can be obtained by rewriting Eqn. (2) as
\begin{displaymath}
\frac{dr_{i}}{dt} = \frac{\mathrm{N}_{LyC}(r_{i})}{4\pi r_{i}^{2}n} = \frac{1}{4\pi nr_{i}^{2}}[\mathrm{N}_{LyC}(0) - \frac{4\pi}{3}r_{i}^{3}n^{2}\alpha^{(2)}],
\end{displaymath}
integration of which yields,
\begin{equation}
r_{i}^{3} = r_{s}^{3}[1 - \mathrm{exp}(-n\alpha^{(2)}t)].
\end{equation}
Maximising this equation gives the timescale, $t_{transform}$, over which the shell of ionising radiation makes a transition from the initial rarefied phase, to the dense phase, which is
\begin{equation}
t_{transform}\sim \frac{1}{n\alpha^{(2)}}.
\end{equation} 
The radius of the expanding shell in the dense phase is given by 
\begin{equation}
\frac{r_{i}(t)}{r_{s}} = \Big(1 + \frac{7}{4}\frac{a_{HII}t}{r_{s}}\Big)^{4/7}.
\end{equation}
(e.g. Shore 2007). For an approximately spherical shell, its average volume density, $\rho_{s}$, is
\begin{equation}
\rho_{s} = \frac{3M_{shell}}{4\pi r_{i}^{3}} = \frac{3M_{shell}}{4\pi r_{s}^{3}(1 + 7a_{HII}t/4r_{s})^{12/7}}.
\end{equation}

After a time $t=t_{transform}$, the shell cools down to an equilibrium temperature, $T_{eq}$, defined by Eqn. (8) below. To estimate the temperature, T$_{eq}$, we shall first account for the likely heating and cooling mechanisms. A crucial contributor towards heating the shell is the photoionisation of H$_{2}$, the corresponding rate of heating is
\begin{equation}
n\Gamma_{i}\sim n^{2}\beta(T_{i})k_{B}T_{i} \ \ \ \mathrm{ergs}\ \mathrm{cm}^{-3}\ \mathrm{s}^{-1};
\end{equation}
(Tielens 2005), where $\beta(T_{i}=10^{4}\mathrm{K}) \sim 1.6\times 10^{-13}$ cm$^{3}$s$^{-1}$, is the recombination cooling coefficient. The cooling rate due to collisional excitation of carbon is
\begin{displaymath}
n^{2}\Lambda_{\mathrm{C}}\sim (3\times 10^{-27})n^{2}\Big(\frac{\mathcal{A}_{C}}{1.4\times 10^{-4}}\Big)\mathrm{exp}\Big(\frac{-92}{T_{i}}\Big) \ \ \ \mathrm{ergs}\ \mathrm{cm}^{-3}\ \mathrm{s}^{-1} 
\end{displaymath}
(Tielens 2005), where $\mathcal{A}_{C}\sim 2.6\times 10^{-4}$, is the abundance of carbon in the ISM.  The equilibrium temperature of the gas, $T_{eq}$, within the shell can be estimated using the condition for thermal equilibrium, $n\Gamma_{i}=n^{2}\Lambda_{C}$, and excess energy will be radiated away so that $(n\Gamma_{i}-n^{2}\Lambda_{C})\sim 3k_{B}T_{eq}/2$, i.e.
\begin{equation}
T_{eq}\sim\frac{2(n\Gamma_{i}-n^{2}\Lambda_{C})}{3k_{B}},
\end{equation}
which is roughly 10 K.

The temperature, $T_{i}$, of the ionised gas is $T_{i}\sim \mathcal{M}^{2}\frac{2\gamma}{(\gamma+1)}T_{ISM}$, where $T_{ISM}=100$ K, is the average temperature of the preshock ISM and $\mathcal{M}$ is the Mach number of the propagating shock, $\mathcal{M}^{2}\sim (v_{i}/a)^{2}$, $a^{2}=\frac{k_{B}T_{ISM}}{\bar{m}_{H}}$. The density of the shocked gas, $\rho_{2}\sim \mathcal{M}^{2}\rho_{1}\sim \mathcal{M}^{2}(n_{H}\bar{m}_{H})$, $\bar{m}_{H}$ is the average mass of atomic hydrogen. Anathpindika (2010), by performing a perturbative analysis, derived an expression for the fastest growing unstable mode in a shocked shell, and the wavenumber, $k$, of this mode is
\begin{equation}
k = \frac{\pi G\Sigma_{s}}{\Big[a_{0}^{2} - \frac{p_{2}}{\rho_{2}}\Big(1 - \frac{r_{out}}{r_{in}}\Big)^{-1}\Big]}.
\end{equation}
Here $r_{out}$ and $r_{in}$ are respectively the outer, and the inner radii of the shell, and $p_{2}/\rho_{2}\sim (dr_{i}/dt)^{2}\equiv V_{s}^{2}$. From Eqn. (3) it follows that,
\begin{equation}
\frac{dr_{i}}{dt} = \frac{r_{s}^{3}}{3}\frac{n\alpha^{(2)}}{r_{i}^{2}}\exp(-n\alpha^{(2)}t).
\end{equation}
For a shell of thickness, $dR_{s}$, $r_{out} = r_{in} + dR_{s}$, so that
$1 - \frac{r_{out}}{r_{in}}\sim \frac{-dR_{s}}{r_{in}}$, and since $M_{shell}\sim 4\pi r_{out}^{2}dR_{s}\rho_{s}\sim 4\pi r_{out}dR_{s}\Sigma_{s}$; $\Sigma_{s}\sim r_{out}\rho_{s}$. For the wavenumber, $k$, calculated using Eqn. (9) above, the corresponding wavelength, $\lambda_{clump} = 2\pi/k$ so that mass of a clump, $M_{clump}\sim \lambda_{clump}^{2}\Sigma_{s}$, and the number of fragments $N_{frag}\sim M_{shell}/M_{clump}$. 
\begin{displaymath}
M_{clump}\sim\lambda_{clump}^{2}\Sigma_{s}\sim \Big(\frac{\pi}{k}\Big)^{2}\Sigma_{s}
\end{displaymath}
\begin{displaymath}
\sim \Big[a_{0}^{2} + V_{s}^{2}\frac{dR_{s}}{r_{in}}\Big]^{2}\frac{1}{G^{2}\Sigma_{s}},
\end{displaymath}
and since $V_{s}>> a_{0}$,
\begin{displaymath}
M_{clump}\sim V_{s}^{4}\Big(\frac{dR_{s}}{r_{in}}\Big)^{2}\frac{1}{G^{2}(r_{out}\rho_{s})}.
\end{displaymath}
The quantity $\Big(\frac{dR_{s}}{r_{in}}\Big)^{2} \sim 10^{-2}$, as will be shown in \S 3 below, so that
\begin{equation}
M_{clump}\sim 10^{3}\Big(\frac{V_{s}}{\mathrm{km/s}}\Big)^{4} \Big(\frac{\mathrm{pc}}{r_{out}}\Big)\Big(\frac{10^{-21}\mathrm{g\ cm}^{-3}}{\rho_{s}}\Big)\mathrm{M}_{\odot}.
\end{equation}
Calculations for typical O-B stars are deferred for \S 3 below.

%----------------------------------------------------------- S_vib
   \begin{figure*}
   \centering
   \includegraphics[angle=270,width=12.cm]{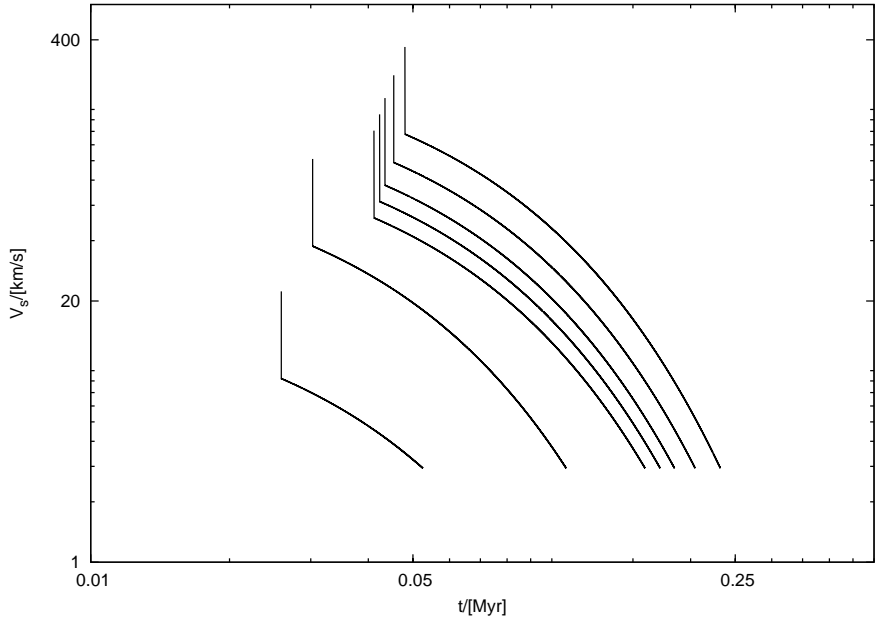}
      \caption{Velocity characteristics for 5 stars belonging to spectral types between O5 and B1 have been plotted. The rightmost characteristic is for a typical O5 star (T$\sim$47000 K \& N$_{LyC}$$\sim 10^{49}$s$^{-1}$), while the leftmost is for a B1 star (T$\sim$22600 K \& N$_{LyC}$$\sim 10^{45}$s$^{-1}$). The transformation of the shell from R-phase to the D-phase occurs at $t\sim t_{transform}$ (Eqn. 4); thereafter the dense shell progressively decelerates. The velocity of the shell, $V_{s}$, in the R-phase, not plotted here, is at least two orders of magnitude higher.
              }
         \label{FigVibStab}
   \end{figure*}
%
%------------------------------------------------------------------

%
%                                                One column figure
%----------------------------------------------------------- S_vib
   \begin{figure*}
   \centering
   \includegraphics[angle=270,width=12.cm]{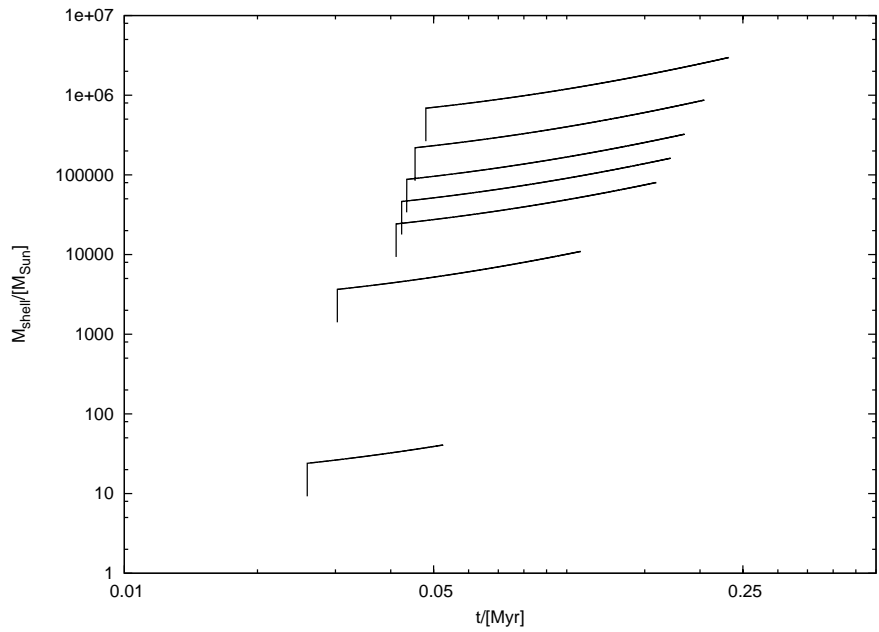}
      \caption{Plots showing the mass swept up by expanding shells for individual stars. The most powerful, O5 star, sweeps up the most massive shell as indicated by the topmost characteristic while the characteristics for stars of other spectral type stack below it. The temperature of driving stars decreases progressively for plots from top to bottom. 
              }
         \label{FigVibStab}
   \end{figure*}
%
%______________________________________________________________

%----------------------------------------------------------- S_vib
   \begin{figure*}
   \centering
   \includegraphics[angle=270,width=12cm]{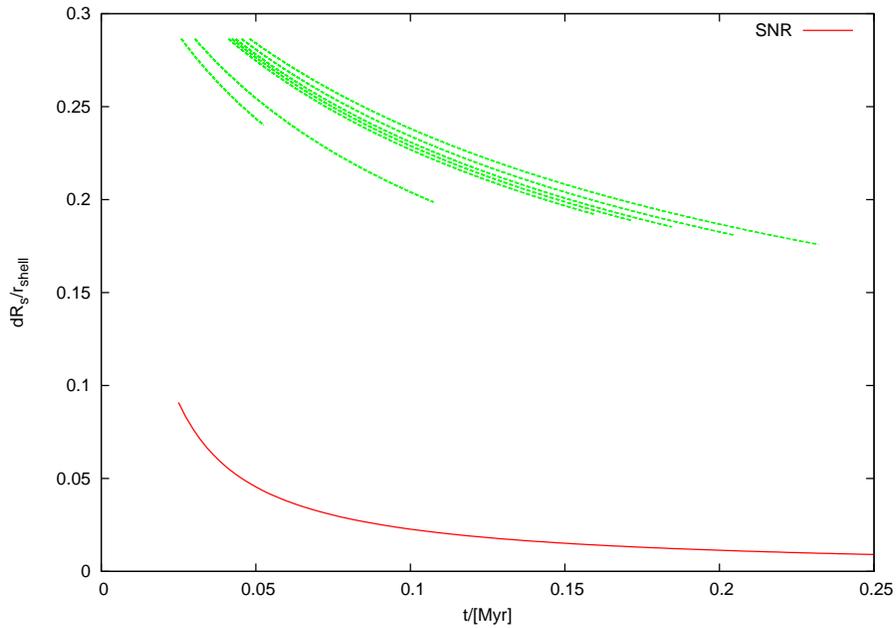}
      \caption{The characteristics plotted in green are for the shells driven by hotter O-type stars. They have similar thickness-to-radius ratio, the curves to the right, extending up to $\gtrsim$0.2 Myrs; while the B-type stars which are much cooler drive comparatively smaller shells and so the ratio is a little higher, at least by 30\%, curves to the left of the former type. The integration for these plots was terminated when respective shells became subsonic. The corresponding ratio for a SNR is much smaller, implying considerably thinner shells.           }
         \label{FigVibStab}
   \end{figure*}

\subsection{Supernova driven shells}

   We now discuss the stability of a shell driven by blast waves originating from a supernova that injects energy, $E_{s}$, into the ISM. For a shell having average density $\rho_{s}$, and radius, $R_{s}$, 
\begin{equation}
E_{s} \sim \frac{4\pi}{3} R_{s}^{4}\rho_{s}\Big(\frac{d^{2}R_{s}}{dt^{2}}\Big).
\end{equation}
 Integrating Eqn. (12) yields the well-known Sedov solution for the shell radius,
\begin{equation} 
R_{s}(t) = C^{\prime}(E_{s}/\rho_{s})^{1/5}t^{2/5},
\end{equation}
where $C^{\prime}$ is a numerical constant of order unity \footnote{$C'\sim (15/4\pi)^{1/5}$}. The post-shock temperature, $T_{i}$, can be calculated using the usual Hugoniot shock conditions for an ISM at a preshock temperature, $T_{ISM}=100$ K, as before. The pressure due to the blast-wave, $p_{b}\sim\rho_{1}k_{B}^{2}T_{i}/\bar{m}_{H}$, under the assumption of approximate isothermality; all symbols have their usual meaning. The average density of the shell, $\rho_{s}\sim\mathcal{M}^{2}\rho_{1}\sim\mathcal{M}^{2}(n_{H}m_{H})$. Using Eqn. (13) above, we can estimate the timescale, $t_{isothermal}$, over which the shell is likely to acquire its equilibrium temperature, $T_{eq}$, defined by Eqn. (8) above, whence it may also slow down substantially, to a sonic or possibly, even sub-sonic speed, $a = \sqrt{k_{B}T_{ISM}/\bar{m}_{H}}$, is the local sound-speed. 

Then at $t=t_{isothermal}$,
\begin{displaymath}
\frac{dR_{s}}{dt} = \Big(\frac{E_{s}}{\rho_{s}}\Big)^{1/5}t_{isothermal}^{-3/5} = \frac{5a}{2}
\end{displaymath}

\begin{equation}
\Rightarrow t_{isothermal} \sim 0.217 a^{-5/3}\Big(\frac{E_{s}}{\rho_{s}}\Big)^{1/3}.
\end{equation}
 The surface density of the shell, $\Sigma_{s}$, is calculated as before and the mass of the shell, $M_{shell}(t=t_{isothermal})$, 
\begin{equation}
M_{shell}\sim 10^{2}\Big(\frac{r_{i}}{\mathrm{pc}}\Big)^{2}\Big(\frac{dR_{s}}{0.1\mathrm{pc}}\Big)\Big(\frac{\rho_{s}}{10^{-21} \mathrm{g cm^{-3}}}\Big) \mathrm{M}_{\odot};
\end{equation}
$r_{i}\equiv r_{s}(t=t_{isothermal})$, is the radius of the shell at that epoch; this expression is also used to calculate the mass of the HII shell. The fastest growing unstable mode, as before, is calculated as before, using Eqn. (9) above. The calculations for a typical supernova remnant (SNR) are demonstrated in the following section.

\section{Results and discussion}
  \emph{The collect and collapse model : The HII shell} \\
For illustrative purposes we consider a typical particle density, $n\sim 10\ \mathrm{cm}^{-3}$. The minimum timescale for radiative cooling of the shell, $t_{transform}$, calculated using Eqn. (4) is approximately 0.1 Myrs for a typical O5 star; we have adopted physical parameters defined in Table 5.3 of Spitzer (1978). The shell, at this epoch, enters in to the dense phase whence it acquires substantial mass during the snow-plough phase, and undergoes substantial deceleration as can be seen from the characteristics plotted in Figs. 1 and 2. Relatively cooler stars of intermediate mass drive weaker ionisation fronts and collect considerably lesser mass. The uppermost characteristic in Fig. 1 shows that a typical O5 star may drive a shell that has a typical mass of a few times $10^{6}$ M$_{\odot}$. An important parameter associated with the stability of an HII shell is its thickness. By a thin shell we imply, $\frac{dR_{s}}{R_{out}}\ll 1$. It can in fact, be easily demonstrated that the ratio of the shell-thickness to radius, may rapidly asymptote to a value less than unity. The average surface density of the shell,
\begin{displaymath}
\Sigma_{s} = \frac{r_{i}^{2}\rho_{s}}{r_{out}},
\end{displaymath}
which following a little manipulation, leads us to
\begin{displaymath} 
\frac{dR_{s}}{r_{i}}\sim\frac{r_{i}\rho_{s}}{\Sigma_{s}}.
\end{displaymath}
For an HII shell, $r_{i}\propto t^{4/7}$, and $\rho_{s}\propto t^{-12/7}$, then for a shell of constant surface density, $\frac{dR_{s}}{r_{i}}\propto t^{-8/7}$. For a typical O5 star, this ratio is $\sim 0.18$. The green curves in Fig. 3 shows time evolution of the HII shell thickness-to-radius ratio for the set of stars used for demonstrative calculations above. These plots suggest, the driven shell is generally thin irrespective of the spectral type of driving star. This is consistent with that reported by the GLIMPSE survey comprising 600 HII shells (Churchwell \emph{et al.} 2006).
Both,  theoretical (e.g Vishniac 1983, 1994), and numerical (e.g. Anathpindika 2009, 2010) work has demonstrated the susceptibility of thin shells to various shearing instabilities, and particularly, to the thin shell instability (TSI). Thin, shock confined shells tend to show a greater proclivity towards the TSI that apparently dominates the classical Jeans instability. The thermal Jeans mass, $M_{Jeans}$, is
\begin{equation}
M_{Jeans}\sim \Big(\frac{a_{0}}{\mathrm{km/s}}\Big)^{3}\Big(\frac{10^{-23}\mathrm{g cm}^{-3}}{\rho_{s}}\Big)^{1/2} 10^{3}\mathrm{M}_{\odot},
\end{equation}
which is roughly an order of magnitude smaller than the typical mass of a clump, $M_{clump}$, defined by Eqn. (11) above which suggests, perturbations triggered by dynamical instabilities in driven shells and/or shocked slabs facilitate concentration of material in perturbed regions, the local maximas or minimas. This was explicitly demonstrated by Wuensch \& Palou$\breve{s}$ (2001). The clump on gaining sufficient mass condenses out, which is the essence of the collect and collapse model. \\

\emph{A supernova shell}

For a demonstrative calculation, let us consider a typical supernova that releases energy, $E_{s}\sim 10^{51}$ ergs, into the ISM that as in the previous case, is assumed to have $n\sim 10$ cm$^{-3}$. The timescale, $t_{isothermal}$, defined by Eqn. (14) above, for a shell of average density $\sim 10^{-21}$ g cm$^{-3}$ is $t_{isothermal}\sim 77$ Myrs, and the velocity of the shell, $V_{s}$, 
\begin{equation}
V_{s} = \frac{dR_{s}}{dt}\Big|_{t=t_{isothermal}},
\end{equation}
which is $\sim 1.5$km/s. The corresponding radius of the shell is then, \\
$R_{s}(t)\sim V_{s}(t)(t=t_{isothermal})\sim 0.11 $ kpc. 
The mass of the shell swept up immediately follows from Eqn. (15), and $M_{shell}\sim 10^{9}$ M$_{\odot}$. Similarly, the mass of a typical fragment calculated using Eqn. (11) is, $M_{clump}\sim 10^{6}$ M$_{\odot}$. It might be interesting to calculate the efficiency, $\delta$ with which the initial energy, $E_{s}$, is converted into mechanical energy whence matter in the ISM is swept up. This efficiency is defined as,
\begin{equation}
\delta = 1 - \Big(\frac{a^{2}_{attenuated}}{a^{2}_{SNR}}\Big)\sim 0.999,
\end{equation} 
which suggests that a large proportion of the initially injected energy is lost in heating the ISM.  The attenuated sound speed, $a_{attenuated}$, defined by Eqn. (19) below accounts for the effects of turbulence generated by various hydrodynamic instabilities which were not directly included in the perturbative analysis that led to Eqn. (9), and
\begin{equation}
a_{attenuated} = \sqrt{\Big[a_{0}^{2} - \frac{p_{2}}{\rho_{2}}\Big(1 - \frac{r_{out}}{r_{in}}\Big)^{-1}\Big]}.
\end{equation}

The GLIMPSE survey of supernovae shells produced a catalogue of 95 SNRs (Reach \emph{et al.} 2006), none of which have shown unambiguous evidence in favour of the collect and collapse model, although a sample of SNRs have shown association with OH masers. However, the source of these masers is unclear and could perhaps be due to sporadic star-formation triggered in molecular clouds over-run by SNRs. Despite this being the case, one may assert that possible condensation in SNRs may take a while, as suggested by the magnitude of $t_{isothermal}$, which is roughly two orders of magnitude larger than the timescale over which the HII shell produced fragments. The resulting fragments are therefore likely to be at least as massive as those forming in the former case.

\subsection{The clump mass spectrum}
The number of clumps, $N_{clump}$, condensing out of a shell is $N_{clump}\sim \frac{M_{shell}}{M_{clump}}$. Then
\begin{displaymath}
\frac{dN_{clump}}{dM_{clump}}\sim -\frac{M_{shell}}{M_{clump}^{2}}dM_{clump} \\
 = -2N_{clump}\Big(\frac{d\lambda}{\lambda}\Big)
\end{displaymath}
or equivalently, in terms of the wavenumber, $k$,
 \begin{equation}
\frac{dN_{clump}}{dM_{clump}} = -2N_{clump}\Big(\frac{dk}{k}\Big).
\end{equation}
This equation defines the number of clumps in an interval $(k,k+dk)$, the integration of which, over the wave number domain yields the mass spectrum for putative clumps. Figure 4 shows a typical mass spectrum for clumps condensing out of HII shells, obtained via a Monte-Carlo integration of Eqn. (20). The integration was performed for 10,000 realisations of shell fragmentation, each producing $N_{clump}$ number of clumps. Thus, we have $\sim$180,000 clumps for the O5 star and fewer, $\sim150,000$, for an O9 star. The relatively smaller shell driven by the latter star not only produces fewer fragments but also, those that are comparatively less massive than those in the former case. However, we note that extremely large clumps, $M_{clump}/M_{\odot}\gtrsim 10^{6}$, may only be sparingly produced as is evident from the mass spectrum in Fig. 4. The mass spectrum for stars of either spectral type is similar and a power-law, $dN/dM_{clump}\propto M_{clump}^{-\beta}$, appears to fit the derived spectrum reasonably well in either cases; $\beta\sim 1.6$, and 1.5 respectively. The spectrum derived here is consistent with the one obtained for large clouds by Fukui \emph{et al.} (1999), and Roslowsky (2005). 

Similarly, we also obtain the mass spectrum for fragments condensing out of the SNR driven shells. The spectrum for this case is plotted in Fig. 5, and a power-law similar to the one in earlier two cases, fits the spectrum reasonably well. In fact, this latter spectrum is similar, $\beta=1.6$, to that for the fragments resulting from the HII shell driven by an O5 star; spectrum for the SNR was derived for $10^{5}$ realisations of shell fragmentation, and shows considerable shift towards a higher mass in comparison to the distributions shown in Fig. 4. This is probably because the velocity of an HII shell in its dense phase decays more rapidly than that of an SNR, which pushes up the attenuated sound-speed, $a_{attenuated}$, for the latter that in turn raises the mass, $M_{clump}$, as seen in Fig. 5.

 Star forming clumps as massive as $\sim$2500 M$_{\odot}$ have been reported in the HII shell N49 and RCW34 (e.g. Zavagno \emph{et al.} 2010, Bik \emph{et al.} 2010). A similar treatment of the problem was presented by Whitworth \emph{et al.} (1994a,b) who arrived at a minimum clump mass of only a few tens of M$_{\odot}$, which in the light of present findings, appears somewhat conservative. 
Our claim more massive clumps is supported by recent observations described above. It is well known, the evolution of a cloud is governed by the dynamical effects associated with the complex interplay between self-gravity, and other contributing factors such as turbulence and the magnetic field that support a clump against the former. It is therefore crucial to predict clump masses with reasonably good accuracy; under-estimation of the mass of putative clumps, for a given radius, which in the present case is roughly equal to the wavelength of the unstable mode, will lower the average density and thus raise the clump-lifetime. An increase in the longevity of clumps will also possibly make events such as clump-clump collisions more probable, than they are known to be. Inter-clump collisions, according to simulations, could lead to bursts of star-formation with fewer low-mass stars (e.g. Chapman  \emph{et al.} 1992, Anathpindika 2009), that will tend to shift the stellar initial mass function towards a top-heavy distribution, away from the widely reported lognormal form.

%----------------------------------------------------------- S_vib
   \begin{figure*}
   \centering
   \includegraphics[angle=270,width=12cm]{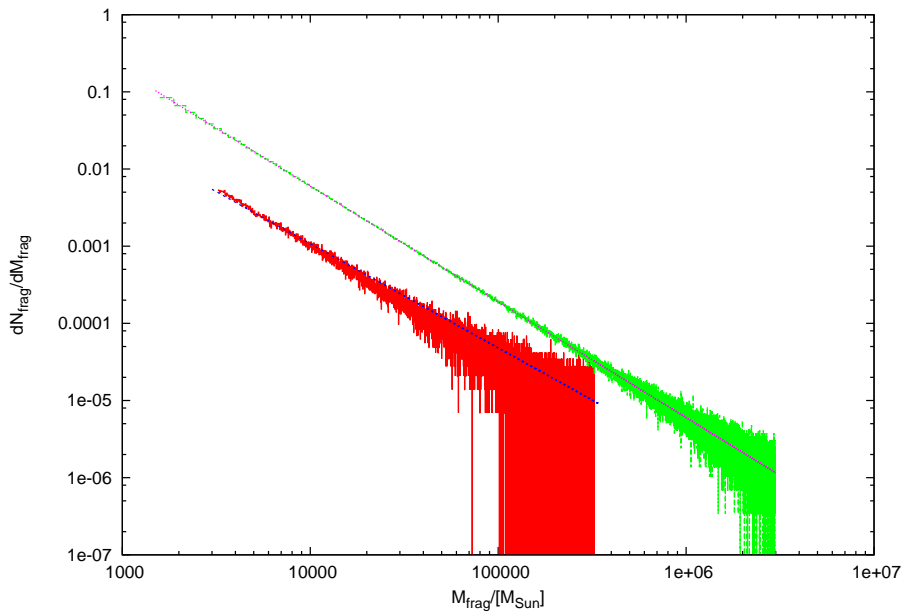}
      \caption{The clump mass spectrum generated via a Monte Carlo integration of Eqn. (20). The green spectrum for a typical O5 star peaks at $\sim10^{3}$ M$_{\odot}$; the peak shifts a little rightward, closer to $\sim 10^{4}$M $_{\odot}$ for an O9 star that is much cooler (red histogram). Formation of massive clumps $\gtrsim 10^{6}$ M$_{\odot}$ is possible, though only sparingly. A power-law fit of the type $\frac{dN}{dM_{clump}}\propto $M$_{clump}^{-\beta}$ agrees reasonably well with the derived spectrum. Also see text below. 
              }
         \label{FigVibStab}
   \end{figure*}
%
%______________________________________________________________

%
%                                                One column figure
%----------------------------------------------------------- S_vib
   \begin{figure*}
   \centering
   \includegraphics[angle=270,width=12.cm]{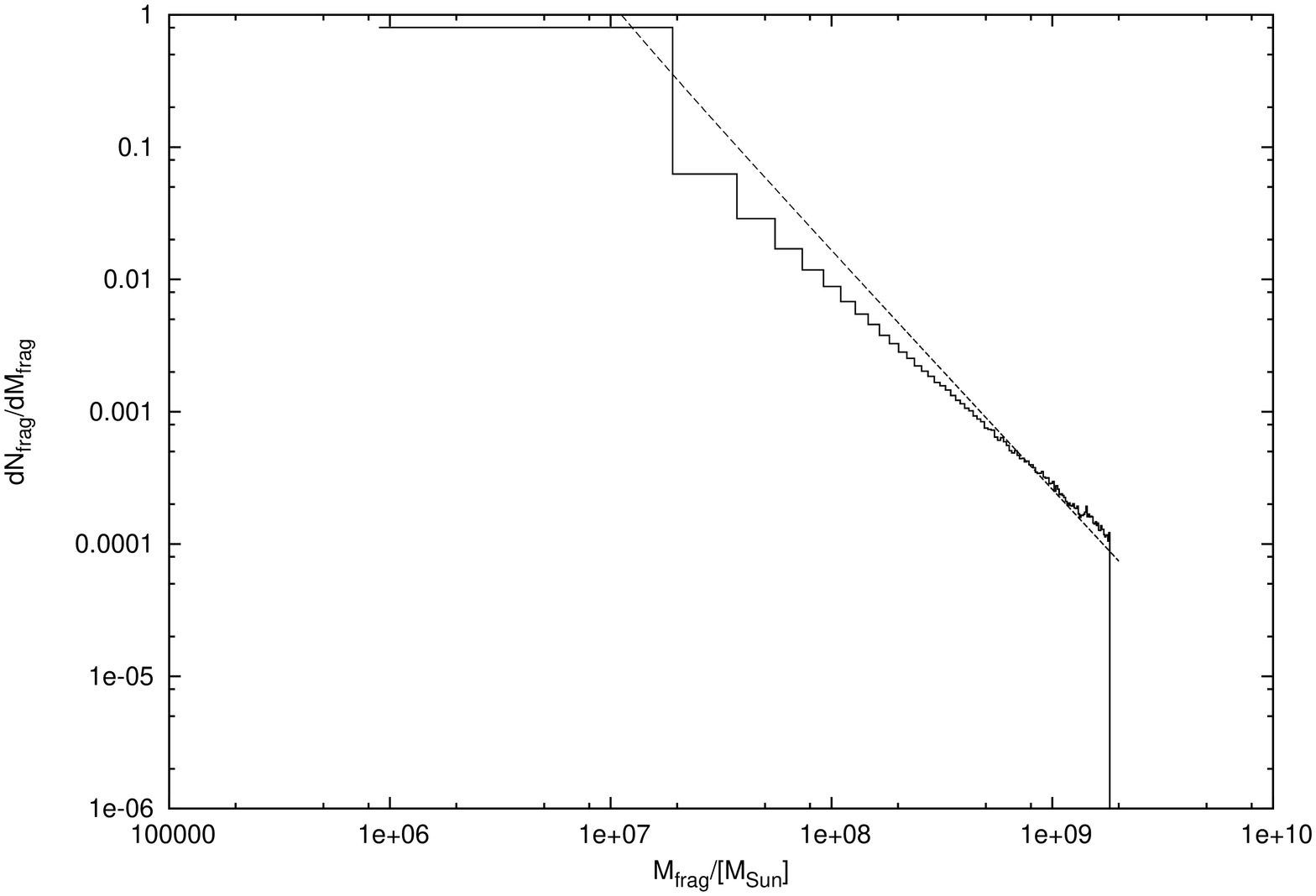}
      \caption{The clump mass spectrum for fragments condensing out of an SNR. The general features of this spectrum are similar to those of the spectrum for clumps produced via fragmentation of a HII shell. See text below.
              }
         \label{FigVibStab}
   \end{figure*}
%
%______________________________________________________________
\section{Conclusions}

We have examined the stability of thin, dense shells driven by powerful ionising radiation originating from massive, early type stars and/or blast waves from a supernova. Our work here shows that fragmentation of these shells is likely to produce large clumps, with masses typically $\gtrsim 10^{3}$ M$_{\odot}$. This calculated range of fragment masses is consistent with that reported via observations of HII shells, and the fragments so formed could spawn a second generation of stars, which may in turn trigger the next generation of stars in the surrounding ISM. The simple calculations discussed above appear to suggest that  the ratio of thickness-to-radius for a shell evolves only weakly with time, and that calculated here (see Fig. 3; $\lesssim 10^{-1}$), is consistent with the values reported in the GLIMPSE survey for HII shells. Similar results are obtained for an SNR, which, however, appears to be much thinner compared to the shells driven by OB stars. This could perhaps be the reason why SNRs often appear filamentary in the IRAC bands (e.g. Reach \emph{et al.} 2006). Simulations studying the evolution of thin shells such those by Anathpindika (2010) have demonstrated the dominance of TSI in shocked shells; the surface of the shell was also shown to develop ripples, similar to breathing modes on fluid surfaces (Fig. 3 in Anathpindika (2010)). While wiggles on the shell surface associated with the TSI though coplanar, are generally orthogonal to its surface, and so in case of a shell in the plane of the sky, cannot alone account for the reported filamentary nature. However, breathing modes coupled with local magnetic field could perchance explain the occurrence of filaments reported by Reach \emph{et al.} (2006).

\emph{\textbf{Acknowledgements}}
      The author is supported by a post doctoral fellowship at the Indian Institute of Astrophysics, Bangalore, India, and wishes to thank an anonymous referee for a critical review of the original manuscript. Useful suggestions from Prof. Harish Bhatt are gratefully acknowledged.

\label{}



%% References


\bibliographystyle{elsarticle-num}
%%\bibliography{<your-bib-database>}

\begin{thebibliography}{00}

 \bibitem{Anathpindika09} Anathpindika, S., 2009,
     A\&A, 504, 437
 \bibitem{Anathpindika10} Anathpindika, S., 2010,
     MNRAS, 405, 1431
 \bibitem{Anderson} Anderson, L., Zavagno, A., Rod{\' o}n, J., \emph{et al.}, 2010, A\&A, 518, L99
 \bibitem{Bik} Bik, A., Puga, E., Waters, L., \emph{et al.}, 2010, 
   ApJ, 713, 883
 \bibitem{Chapman} Chapman, S., Pongracic, H., Disney, M., Nelson, A., Turner, J \& Whitworth, A., 1992, Nature, 359, 207
 \bibitem{Churchwell} Churchwell, E., Povich, M., Allen, D., \emph{et al.}, 2006, ApJ, 649, 759
 \bibitem{Deharveng} Deharveng,L., Zavagno,A., Caplan, J., 2005,
      A\&A, 433, 565
 \bibitem{Ehlerova}Ehlerov$\grave{a}$, S \& Palou$\breve{s}$, J., 2002, MNRAS, 330, 1026
 \bibitem{Elmegreen} Elmegreen, B \& Elmegreen, D., 1978, ApJ, 220, 1051  
 \bibitem{Elmegreen} Elmegreen, B., 1989, ApJ, 344, 306
 \bibitem{Fukui} Fukui, Y., Onishi, T., Abe, R., Kawamura, A., Tachihara, K., Yamaguchi, R., Mizuno, A \& Ogawa, H., 1999, PASJ, 51, 751
 \bibitem{Larson}Larson, R., 1985, MNRAS, 214, 379
 \bibitem{Reach}Reach, W., Rho, J. \emph{et al.}, 2006, ApJ, 131, 1479
 \bibitem{Roslowsky}Roslowsky, E., 2005, PASJ, 117, 1403
 \bibitem{Spitzer}Spitzer, L., 1978, in \emph{Physical Processes in Interstellar medium}, Wiley-Intersceince. Pub.
 \bibitem{Shore}Shore, S., \emph{Astrophysical Hydrodynamics}, Wiley-VCH, Weinheim, 2007; p. 141-43 
 \bibitem{Tielens}Tielens, A., G., 2005, \emph{The physics and chemistry of the ISM}, Cambridge university Press, UK, p. 54-6
 \bibitem{Vishniac} Vishniac, E., 1983, ApJ, 274, 152
 \bibitem{Vishniac} Vishniac, E., 1994, ApJ, 428, 186
 \bibitem{Whitwortha} Whitworth, A., Bhattal, A., Chapman, S., Disney, M \& Turner, J., 1994a, MNRAS, 268, 291
 \bibitem{Whitworthb} Whitworth, A., Bhattal, A., Chapman, S., Disney, M \& Turner, J., 1994b, A\&A, 290, 421
 \bibitem{Wuensch}Wuench, R \& Palou$\breve{s}$, J., 2001, A\&A, 374, 746
 \bibitem{Yamaguchi}Yamaguchi, R., Norikazu, M., Onishi, T., Mizuno, A \& Fukui, Y., 2001, ApJ, 553, L185
 \bibitem{zavagno} Zavagno, A., Deharveng, L., Comer{\' o}n, F., Brand, J., Massy, F., Caplan, J \& Russeil, D., 2006, A\&A, 171, 184
 \bibitem{Zavagno} Zavagno, A., Anderson, L., Russeil, D., \emph{et al.}, 2010,
    A\&A, 518, L101
     



 \end{thebibliography}



\end{document}